\documentclass[aps,prc,
superscriptaddress,
preprint,
showpacs]{revtex4-1}
\usepackage{amsmath}
\usepackage{color}
\usepackage{graphicx}

\begin{document}
\title{Investigation of ${}^{10}$Be and its cluster dynamics from
  nonlocalized clustering concept}

\author{Mengjiao Lyu}
\email{mengjiao\_lyu@hotmail.com.}  \affiliation{School of Physics and
  Key Laboratory of Modern Acoustics, Institute of Acoustics, Nanjing
  University, Nanjing 210093, China} \affiliation{Research Center for
  Nuclear Physics (RCNP), Osaka University, Osaka 567-0047, Japan}

\author{Zhongzhou Ren} \email{zren@nju.edu.cn.}  \affiliation{School
  of Physics and Key Laboratory of Modern Acoustics, Institute of
  Acoustics, Nanjing University, Nanjing 210093, China}
\affiliation{Center of Theoretical Nuclear Physics, National
  Laboratory of Heavy-Ion Accelerator, Lanzhou 730000, China}

\author{Bo Zhou} \email{bo@nucl.sci.hokudai.ac.jp}
\affiliation{Faculty of Science, Hokkaido University, Sapporo
  060-0810, Japan}

\author{Yasuro Funaki} \affiliation{Nishina Center for
  Accelerator-Based Science, The Institute of Physical and Chemical
  Research (RIKEN), Wako 351-0198, Japan}

\author{Hisashi Horiuchi} \affiliation {Research Center for Nuclear
  Physics (RCNP), Osaka University, Osaka 567-0047, Japan}
\affiliation {International Institute for Advanced Studies, Kizugawa
  619-0225, Japan}

\author{\mbox{Gerd R\"{o}pke}} \affiliation{Institut f\"{u}r Physik,
  Universit\"{a}t Rostock, D-18051 Rostock, Germany}

\author{Peter Schuck} \affiliation{Institut de Physique Nucl\'{e}aire,
  Universit\'e Paris-Sud, IN2P3-CNRS, UMR 8608, F-91406, Orsay,
  France} \affiliation{Laboratoire de Physique et Mod\'elisation des
  Milieux Condens\'es, CNRS-UMR 5493, F-38042 Grenoble Cedex 9,
  France}

\author{Akihiro Tohsaki} \affiliation{Research Center for Nuclear
  Physics (RCNP), Osaka University, Osaka 567-0047, Japan}

\author{Chang Xu} \affiliation{School of Physics and Key Laboratory of
  Modern Acoustics, Institute of Acoustics, Nanjing University,
  Nanjing 210093, China}

\author{Taiichi Yamada} \affiliation{Laboratory of Physics, Kanto
  Gakuin University, Yokohama 236-8501, Japan}

\date{\today}

\begin{abstract}
  We extend the new concept of nonlocalized clustering to the
  nucleus $^{10}$Be with proton number Z=4 and neutron number N=6 (N=Z+2).
  The Tohsaki-Horiuchi-Schuck-R\"{o}pke (THSR) wave
  function is formulated for the description of different 
  structures of $^{10}$Be.
  Physical properties such as energy spectrum and 
  root-mean-square radii are calculated for the first two $0^{+}$
  states and corresponding rotational bands.  With only one single
  THSR wave function, the calculated results show good agreement with
  other models and experimental values. We apply, for the first time,
  the THSR wave function on the chain orbit ($\sigma$-orbit) structure
  in the $0_{2}^{+}$ state of $^{10}$Be. The ring orbit ($\pi$-orbit)
  and $\sigma$-orbit structures are further illustrated by calculating
  the density distribution of the valence neutrons. We also
  investigate the dynamics of $\alpha$-clusters and the correlations of
  two valence neutrons in ${}^{10}$Be.
\end{abstract}


\maketitle

\section{Introduction}
Cluster formation plays a fundamental role in understanding
the structure and properties of nuclei.
In recent years, tremendous progress has been made in the
investigation of cluster structure in light nuclei \cite{Tohsaki2001,
  Funaki2002, Yamada2005, Zhou2012, Zhou2013, Zhou2014, Suhara2014,
  Lyu2015, Enyo1995, Xu2006, Yren2012, Ma2014, Ye2014}, especially due to the
model wave function with nonlocalized clustering concept, namely the
THSR wave function \cite{Tohsaki2001, Funaki2002, Yamada2005,
  Zhou2012, Suhara2014, Zhou2013, Zhou2014, Lyu2015}. The THSR wave
function was first proposed to describe the $\alpha$-cluster
condensation in gas-like states, including the famous Hoyle state
($0_{2}^{+}$ state) in ${}^{12}$C \cite{Tohsaki2001}. Then it was
successfully applied to various other aggregates of $\alpha$-clusters
such as ${}^{8}$Be, ${}^{16}$O, ${}^{20}$Ne \cite{Funaki2002,
  Tohsaki2001, Zhou2012}, and also to one-dimensional chain
systems \cite{Suhara2014}.  It was found that one single THSR wave
function is almost 100\% equivalent to the RGM/GCM (Resonating Group
Method/Generator Coordinate Method) wave functions for both gas-like
and non-gaslike states.  In the recent studies of
inversion-doublet-bands of ${}^{20}$Ne \cite{Zhou2013, Zhou2014}, the
nonlocalized character of clustering rooted in the THSR wave function
is proved to be a very important property for clustering structure in
light nuclei. Therefore it is very necessary to investigate the
nonlocalized cluster dynamics in other different nuclear systems.

In recent years, there are also investigations using the THSR wave
function and its intrinsic container structure for nuclei beyond
traditional $\alpha$ aggregates.  This starts with the calculation of
${}^{13}$C with a neutron probe interacting with the
3$\alpha$-condensation \cite{Tohsaki2008}.  Also, in this year, the
2$\alpha$+$\Lambda$ system is investigated with Hyper-THSR wave
function, which shows the essential role of the container picture for
the cluster structure in $^{9}_{\Lambda}$Be \cite{Funaki2015}.  In our
previous work, the THSR wave function is constructed with intrinsic
negative parity and applied in the calculation of nucleus ${}^{9}$Be
with proton number Z=4 and neutron number N=5 (N=Z+1) \cite{Lyu2015}. In this
study, the nonlocalized clustering concept is shown to prevail in the
$\pi$-orbit for ${}^{9}$Be.

In order to apply the nonlocalized clustering concept to more general
nuclei, it is very interesting to extend the THSR wave function to the
investigation of the N=Z+2 cluster nucleus $^{10}$Be in which one more
valence neutron is added to the $^{9}$Be system. The nucleus $^{10}$Be
is well-known for its typical nuclear molecular orbit structure, and
has been studied with many different models \cite{Oertzen1996,
  Itagaki2000, Ito2004, Warburton1992, Kobayashi2011, Kobayashi2012,
  Itagaki2008, Myo2015}. The $0_{2}^{+}$ state of $^{10}$Be is
considered to be an intruder state which is difficult to be described
by simple shell model methods \cite{Warburton1992}. Besides, a chain
structure with $\sigma$-binding and enormous spatial extension is
found in this $0_{2}^{+}$ state \cite{Itagaki2000, Ito2004}. Another
interesting topic for the N=Z+2 nucleus $^{10}$Be is the correlation
between valence neutrons \cite{Kobayashi2011}. Recently, predictions
of the $0_{3}^{+}$ and $0_{4}^{+}$ states \cite{Kobayashi2012} and the
existence of $\alpha$+t+t structure in $^{10}$Be \cite{Itagaki2008}
have been reported.  In the present work, we construct THSR wave
functions for $^{10}$Be based on the new nonlocalized picture. With
these THSR wave functions, we can well describe not only the physical
properties of different states of $^{10}$Be but also the cluster
dynamics in these states with only one single THSR wave function.

We organize this paper as follows. In Section II
we formulate the THSR wave function for both $\pi$- and
$\sigma$-binding of $^{10}$Be.  In Section III, we present our
results for the $0_{1}^{+}$ ground state of $^{10}$Be and its
rotational bands. In Section IV, we investigate the $0_{2}^{+}$
state of $^{10}$Be and its $\sigma$-orbit structure. In Section V, we
discuss the $\alpha$-cluster dynamics and correlations between valence
neutrons in these states. The last Section VI contains the conclusions.

\section{Formulation of the THSR Wave Function for
  ${}^{10}$B\MakeLowercase{e}}
\label{sec:waveFunction}
\begin{figure}[htbp]
  \centering
  \includegraphics[width=0.8\textwidth]{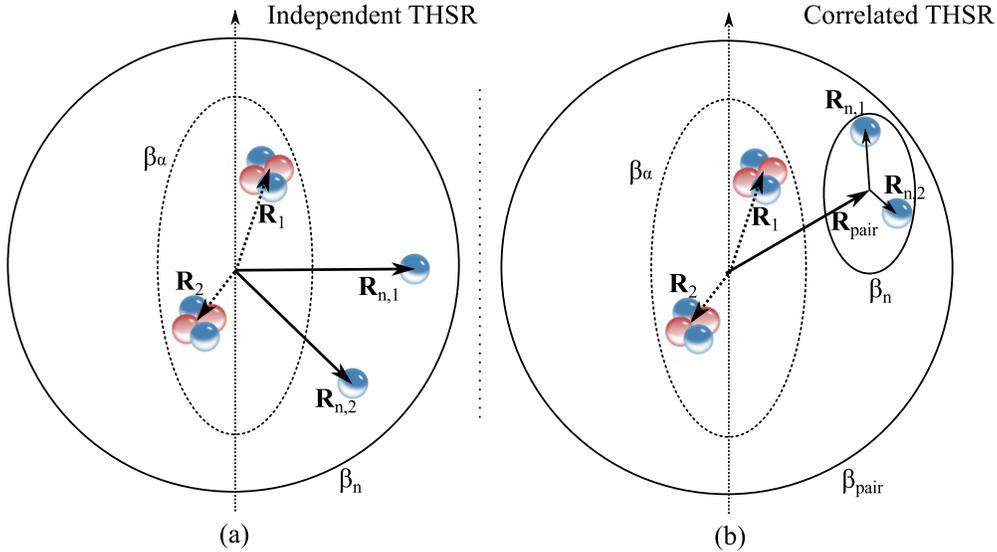}
  \caption{\label{fig:container}(Color online.) Generator coordinates and $\beta$
    parameters used in the THSR wave function for the ground state
    $0_{1}^{+}$ of $^{10}$Be. Left panel (a) shows the case without
    correlations between valence neutrons. Right panel (b) shows the
    case of correlated valence neutron motion. Vectors are
    corresponding generator coordinates. Dashed ellipses denote
    nonlocalized motion of $\alpha$-clusters. Solid ellipses denote
    the nonlocalized motion of the valence neutrons or of the correlated
    two neutron-valence sub-system.}
\end{figure}

We first introduce the THSR wave function of $^{10}$Be for the
$\pi$-orbit binding structure of the two valence neutrons without considering their 
correlations, as shown in the left panel of Fig.~\ref{fig:container}.
This wave function, designated as the independent THSR wave function
$\left| \Phi_{\textrm{ind}} \right\rangle$, can simply be constructed with the
same form as used for $^{9}$Be in our previous work \cite{Lyu2015},
\begin{equation}
\label{eq:indwf}
  \left| \Phi_{\textrm{ind}} ({}^{10}\text{Be}) \right\rangle
  =(C_{\alpha}^{\dagger})^{2}
   c_{n,\uparrow}^{\dagger}
   c_{n,\downarrow}^{\dagger}\left| {\textrm{vac}} \right\rangle,
\end{equation}
where $C_{\alpha}^{\dagger}$ and $c_{n}^{\dagger}$ are creation
operators of $\alpha$-clusters and valence neutrons, respectively.
The $\alpha$-creator $C_{\alpha}^{\dagger}$ determines the dynamics of the
$\alpha$-clusters and can be written as
\begin{equation}
  \label{eq:alphaCreator}
  \begin{split}
  C_{\alpha}^{\dagger}=\int &d^3\mathbf{R}
    \exp (-\frac{R_{x}^{2}}{\beta_{\alpha,xy}^{2}}
    -\frac{R_{y}^{2}}{\beta_{\alpha,xy}^{2}}
    -\frac{R_{z}^{2}}{\beta_{\alpha,z}^{2}})\int d^{3}\mathbf{r}_{1}
      \cdots d^{3}\mathbf{r}_{4}    \\
  &\times \psi(\mathbf{r}_{1}-\mathbf{R})
      a_{\sigma_{1},\tau_{1}}^{\dagger}(\mathbf{r}_{1})
    \cdots \psi(\mathbf{r}_{4}-\mathbf{R})
      a_{\sigma_{4},\tau_{4}}^{\dagger}(\mathbf{r}_{4}),
  \end{split}
\end{equation}
where $\mathbf{R}$ is the generate coordinate of the $\alpha$-cluster,
$\mathbf{r}_i$ is the position of the $i$th nucleon.
$a_{\sigma,\tau}^{\dagger}(\mathbf{r}_i)$ is the creation operator of
the $i$th nucleon with spin $\sigma$ and isospin $\tau$ at position
$\mathbf{r}_i$.
$\psi(\mathbf{r}) =(\pi b^{2})^{-3/4} \exp(-r^{2}/2b^{2})$ is the wave
function of a single nucleon in the $\alpha$-clusters with a Gaussian
form where the parameter $b$ in this Gaussian describes the size of
$\alpha$-clusters. $\beta_{\alpha,xy}$ and $\beta_{\alpha,z}$ are
parameters for the nonlocalized motion of two $\alpha$-clusters in
horizontal or vertical directions respectively, which is shown as a
dashed ellipse in the left panel of Fig.~\ref{fig:container}.  For the
valence neutrons, we use the same creation operator $c_{n}^{\dagger}$
as in our previous study of $^{9}$Be \cite{Lyu2015}
\begin{equation}
  \label{eq:extraCreator}
  \begin{split}
  c_{n,\sigma}^{\dagger}=\int &d^{3}\mathbf{R}_{n}
    \exp (-\frac{R_{n,x}^{2}}{\beta_{n,xy}^{2}}
          -\frac{R_{n,y}^{2}}{\beta_{n,xy}^{2}}
          -\frac{R_{n,z}^{2}}{\beta_{n,z}^{2}})
    e^{im\phi_{\mathbf{R}_{n}}}\int d^{3}\mathbf{r}_n    \\
  &\times (\pi b^{2})^{-3/4}
  e^{-\frac{(\mathbf{r}_n-\mathbf{R}_{n})^{2}}{2b^{2}}}
    a_{\sigma,n}^{\dagger}(\mathbf{r}_n),
  \end{split}
\end{equation}
where $\mathbf{R}_{n}$ is the generate coordinator of the valence neutrons,
$\mathbf{r}_n$ is the position of the extra neutron,
$a_{\sigma,\tau}^{\dagger}(\mathbf{r}_n)$ is the creation operator of
the extra neutron with spin $\sigma$ at position $\mathbf{r}_n$, and
$\phi_{\mathbf{R}_{n}}$ is the azimuthal angle in spherical
coordinates
$(R_{\mathbf{R}_{n}}, \theta_{\mathbf{R}_{n}}, \phi_{\mathbf{R}_{n}})$
of $\mathbf{R}_{n}$. In this creation operator, the size parameter $b$
of the Gaussian is taken to be  the same as that in
Eq.~(\ref{eq:alphaCreator}).  $\beta_{n,xy}$ and $\beta_{n,z}$ are
parameters for the nonlocalized motion of the extra neutron, shown as
solid ellipse in the left panel of Fig.~\ref{fig:container}.

In Eq.~(\ref{eq:indwf}), the valence neutrons are assumed to move freely
and independently of each other. However, dineutron correlations play
an important role for nuclei with two valence nucleons such as $^{6}$He
and $^{10}$Be, as discussed previously in Ref.~\cite{Kobayashi2011}.
The introduction of correlations in the THSR wave function is very natural,
because the original THSR wave function of $\alpha$-aggregates
describes very well the $\alpha$-$\alpha$ correlations, especially in
Refs.~\cite{Zhou2013, Zhou2014}. Thus, we can describe the correlation
of two valence neutrons in the two-neutron sub-system with a new THSR
integration analogously to what we used for two $\alpha$-clusters.

As shown in the right panel of Fig.~\ref{fig:container}, two valence
neutrons can form a correlated sub-system. The relative position of the valence
neutrons in this sub-system is labeled by the generator coordinate
$\mathbf{R}_{n}$, while the position of the correlated sub-system can
be described by the generator coordinate $\mathbf{R}_{{\textrm{pair}}}$. Thus the
generator coordinate for the valence neutrons should be
$\mathbf{R}_{{\textrm{pair}}}+\mathbf{R}_{n}$.  With these generator coordinates,
we can write the THSR wave function with dineutron correlation in
$^{10}$Be as
\begin{equation}
\label{eq:corrwf}
\begin{split}
         \left| \Phi_{\textrm{cor}} ({}^{10}\text{Be})\right\rangle
         &=(C_{\alpha}^{\dagger})^{2}c_{{\textrm{pair}}}^{\dagger}\left| {\textrm
             vac} \right\rangle,
\end{split}
\end{equation}
where $ c_{{\textrm{pair}}}^{\dagger} $ is the creation operator for the dineutron
pair, which can be denoted as
\begin{equation}
\label{eq:pairCreator}
\begin{split}
	c_{{\textrm{pair}}}^{\dagger}=\int &d^{3}\boldsymbol{R}_{{\textrm{pair}}}
	\exp (-\frac{R_{{\textrm{pair}},x}^{2}}{\beta_{{\textrm{pair}},xy}^{2}}
	-\frac{R_{{\textrm{pair}},y}^{2}}{\beta_{{\textrm{pair}},xy}^{2}}
	-\frac{R_{{\textrm{pair}},z}^{2}}{\beta_{{\textrm{pair}},z}^{2}})
          c_{n,\uparrow}^{\dagger}(\boldsymbol{R}_{{\textrm{pair}}})
          c_{n,\downarrow}^{\dagger}(\boldsymbol{R}_{{\textrm{pair}}}).
\end{split}
\end{equation}
This integration of $\mathbf{R}_{{\textrm{pair}}}$ determines the collective
motion of the two-neutron sub-system.  $\beta_{{\textrm{pair}},xy}$ and
$\beta_{{\textrm{pair}},z}$ are parameters for this collective motion, which is
shown as the big solid ellipse in the right panel of
Fig.~\ref{fig:container}.
$c_{n, \sigma}^{\dagger} (\boldsymbol{R}_{{\textrm{pair}}})$ is the creation
operator for each neutron, which has a similar form as
$c_{n, \sigma}^{\dagger}$ in Eq.~(\ref{eq:extraCreator}),
\begin{equation}
\label{eq:pairNeutronCreator}
\begin{split}
c_{n,\sigma}^{\dagger}(\boldsymbol{R}_{{\textrm{pair}}})=&
  \int d^{3}\boldsymbol{R}_{n}
        \exp (-\frac{R_{n,x}^{2}}{\beta_{n,xy}^{2}}
	-\frac{R_{n,y}^{2}}{\beta_{n,xy}^{2}}
	-\frac{R_{n,z}^{2}}{\beta_{n,z}^{2}})
	{e^{im\phi(\boldsymbol{R}_{{\textrm{pair}}}+\boldsymbol{R}_{n})}}\\
&\times
	\int d^{3}\boldsymbol{r} (\pi b^{2})^{-3/4}
	e^{-\frac{(\boldsymbol{r}-\boldsymbol{R}_{{\textrm{pair}}}-\boldsymbol{R}_{n})^{2}}
                  {2b^{2}}}
	a_{\sigma,n}^{\dagger}(\boldsymbol{r}).\\
\end{split}
\end{equation}
This integration of $\mathbf{R}_{n}$ determines the relative motion of
two valence neutrons inside the two-neutron sub-system.
$\beta_{n,xy}$ and $\beta_{n,z}$ are parameters for this relative
motion, which is shown as small solid ellipse in the right panel of
Fig.~\ref{fig:container}.

As illustrated in Ref.~\cite{Lyu2015}, the phase factor
$e^{im\phi_{\mathbf{R}_{n}}}$ in Eq.~(\ref{eq:extraCreator}) with
parameter $m=\pm 1$ ensures negative parity for the single nucleon wave
function of the valence neutron. The same argument can also be
applied to the phase factor
$e^{im\phi(\boldsymbol{R}_{{\textrm{pair}}} +\boldsymbol{R}_{n})}$ in
Eq.~(\ref{eq:pairNeutronCreator}).  Besides, when $m=0$, only Gaussian
functions are left in the creation operators in Eq.~(\ref{eq:extraCreator})
or Eq.~(\ref{eq:pairNeutronCreator}), what corresponds to a positive
parity for the valence neutrons. Therefore, we have the total parity
of ${}^{10}$Be as
\begin{equation}\label{eq:parity}
  \pi=\pi_{\alpha}^{(1)}\times\pi_{\alpha}^{(2)}
      \times\pi_{n}^{(1)}\times\pi_{n}^{(2)}=
  \begin{cases}
    (+)\times(+)\times(+)\times(+)=+& (m_{1}=m_{2}=0)\\
    (+)\times(+)\times(-)\times(-)=+& (m_{1}=1,m_{2}=-1).
  \end{cases}
\end{equation}

\begin{figure}[htbp]
  \centering
  \includegraphics[width=0.45\textwidth]{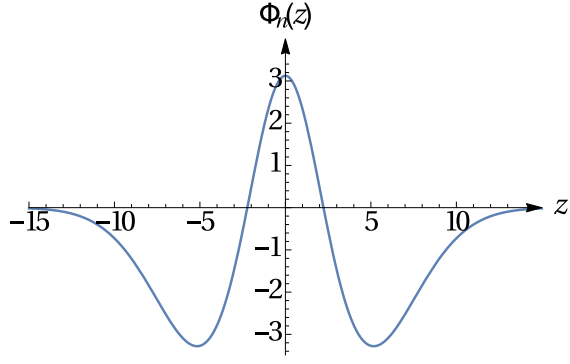}
  \caption{\label{fig:sigma-fz}Single nucleon wave function $\phi_{n}(z)$
    for the valence neutron in the $0_{2}^{+}$ state of
    $^{10}$Be. Parameters are chosen as $D=2$ fm and $\beta_{n,z}=5$
    fm.  }
\end{figure}

For the $0_{2}^{+}$ state of $^{10}$Be, it is already known that this
state has a very typical chain structure because of the $\sigma$-binding
mechanism \cite{Ogawa2000, Itagaki2000}. In this structure, valence
neutrons stay between or outside of two $\alpha$-clusters along the
$\alpha$-$\alpha$ chain. For this state, we construct the
one-dimensional constrained THSR wave function for independent neutrons
\begin{equation}
\label{eq:chainWF}
  \left| \Phi_{\textrm{chain}} ({}^{10}\text{Be}) \right\rangle
  =(C_{\alpha}^{\dagger})^{2}
   c_{n,\uparrow}^{\dagger}
   c_{n,\downarrow}^{\dagger}\left| {\textrm vac} \right\rangle.
\end{equation}
Here the $\alpha$-creation operator $C_{\alpha}^{\dagger}$ is similar to
Eq.~(\ref{eq:alphaCreator}) but operates only on the $z$-axis,
\begin{equation}
  \label{eq:alphaCreatorSigma}
  \begin{split}
  C_{\alpha}^{\dagger}=\int &d R_{z}
    \exp (-\frac{R_{z}^{2}}{\beta_{\alpha,z}^{2}})
    \int d^{3}\mathbf{r}_{1}
      \cdots d^{3}\mathbf{r}_{4}    \\
  &\times \psi(\mathbf{r}_{1}-\mathbf{R})
      a_{\sigma_{1},\tau_{1}}^{\dagger}(\mathbf{r}_{1})
    \cdots \psi(\mathbf{r}_{4}-\mathbf{R})
      a_{\sigma_{4},\tau_{4}}^{\dagger}(\mathbf{r}_{4}),
  \end{split}
\end{equation}
where $\mathbf{R}$ is the generate coordinate of the $\alpha$-cluster
on the $z$-axis and $\mathbf{r}_i$ is the position of the $i$th
nucleon. For the valence neutrons, a creation operator with a node
structure is constructed for the correct description of the
$\sigma$-orbits, as
\begin{equation}
  \label{eq:extraCreatorSigma}
  \begin{split}
  c_{n,\sigma}^{\dagger}=\int &d R_{n,z}
    (D-|R_{n,z}|)\exp (
          -\frac{R_{n,z}^{2}}{\beta_{n,z}^{2}})
    \int d^{3}\mathbf{r}_n    \\
  &\times (\pi b^{2})^{-3/4}
  e^{-\frac{(\mathbf{r}_n-\mathbf{R}_{n})^{2}}{2b^{2}}}
    a_{\sigma,n}^{\dagger}(\mathbf{r}_n).
  \end{split}
\end{equation}
Here, $\mathbf{R}_{n}$ is the generate coordinate of the valence
neutron on the $z$-axis and $\mathbf{r}_{n}$ is the position of the
valence neutron. We introduce a new factor $(D-|R_{n,z}|)$ in this
THSR integration to provide a node structure for the wave function of
the valence neutrons. Two nodes appear in the wave function when
$R_{n,z}=\pm D$, which are locations near the $\alpha$-clusters.  As a
demonstration, we choose parameter $D=2$ fm and $\beta_{n,z}=5$ fm,
and show the single nucleon wave function $\phi_{n}(z)$ for valence
neutron in Fig.~\ref{fig:sigma-fz}. It is clearly seen in the figure that
the wave function $\phi_{n}(z)$ has a different sign between the
mid-region and two flanks. Also, two nodes appear near $z=2$ fm as
expected. In our calculation, the parameter $D$ is treated as a
variational parameter to obtain a correct node position for the wave
function $\phi_{n}(z)$.

It is obvious that Eq.~(\ref{eq:alphaCreatorSigma}) and
Eq.~(\ref{eq:extraCreatorSigma}) contain only even functions, so the
corresponding parity in this wave function is given by,
\begin{equation}\label{eq:paritySigma}
  \pi=\pi_{\alpha}^{(1)}\times\pi_{\alpha}^{(2)}
      \times\pi_{n}^{(1)}\times\pi_{n}^{(2)}=
    (+)\times(+)\times(+)\times(+)=+.
\end{equation}

In order to eliminate effects from spurious center-of-mass (c.o.m.)
motion, the c.o.m. part of $\left| \Phi \right\rangle$ is projected
onto a $(0s)$ state \cite{Okabe1977}. We use the following
transformation of coordinates $\mathbf{r}_{i}$ in
$\left| \Phi \right\rangle$ to eliminate the effects of the spurious
center-of-mass motion as in Ref.~\cite{Okabe1977}
\begin{equation}
\left| \Psi \right\rangle=\left| (0s)\text{c.o.m.} \right\rangle\rangle
    \left\langle\langle (0s)\text{c.o.m.}  | \Phi \right\rangle.
\end{equation}
Here $(0s)$ represents the wave function of the c.o.m. coordinate
$\mathbf{X}_{G}$ in $s$-state, and the double brackets denotes the
  integration with respect to coordinate $\mathbf{X}_{G}$.  We also apply the
angular-momentum projection technique
$\hat{P}_{MK}^{J}\left| \Psi \right\rangle$ to restore the rotational
symmetry \cite{Schuck1980},
\begin{equation}\label{eq:ap}
  \begin{split}
\left| \Psi^{JM} \right\rangle&=\hat{P}_{MK}^{J}\left| \Psi \right\rangle\\
    &=\frac{2J+1}{8\pi^{2}}\int d \Omega D^{J*}_{MK}(\Omega)\hat R (\Omega)
    \left| \Psi \right\rangle,
  \end{split}
\end{equation}
 where $J$ is the total angular momentum of ${}^{10}$Be.

The Hamiltonian of the ${}^{10}$Be system can be written as
\begin{equation}\label{hamiltonian}
  H=\sum_{i=1}^{10} T_i-T_{c.m.} +\sum_{i<j}^{10}V^N_{ij}
    +\sum_{i<j}^{10}V^C_{ij} +\sum_{i<j}^{10}V^{ls}_{ij},
\end{equation}
where $T_{c.m.}$ is the kinetic energy of the center-of-mass
motion. Volkov No. 2 \cite{Volkov1965} is used as the central force of
the nucleon-nucleon potential,
\begin{equation}\label{vn}
  V^N_{ij}=\{V_1 e^{-\alpha_1 r^2_{ij}}-V_2 e^{-\alpha_2 r^2_{ij}}\}
  \{ W - M \hat P_\sigma \hat P_\tau \ + B \hat P_{\sigma} - H \hat P_{\tau}\},
\end{equation}
where $M=0.6$, $W=0.4$ and $B=H=0.125$. Other parameters are
$V_{1}=-60.650$ MeV, $V_{2}=61.140$ MeV, $\alpha_{1}=0.309$
fm${}^{-2}$, and $\alpha_{2}=0.980$ fm${}^{-2}$.  The G3RS (Gaussian
soft core potential with three ranges) term is taken as the two-body
type spin-orbit interaction \cite{Yamaguchi1979},
\begin{equation}\label{vc}
V^{ls}_{ij}=V^{ls}_0\{
           e^{-\alpha_1 r^2_{ij}}-e^{-\alpha_2 r^2_{ij}}
           \} \mathbf{L}\cdot\mathbf{S} \hat{P}_{31},
\end{equation}
where $\hat P_{31}$ projects the two-body system onto triplet odd
state. Parameters in $V^{ls}_{ij}$ are taken from
Ref.~\cite{Kobayashi2012} with $V_{0}^{ls}$=1600 MeV,
$\alpha_{1}$=5.00 fm${}^{-2}$, and $\alpha_{2}$=2.778 fm${}^{-2}$.

\section{The $0_{1}^{+}$ ground state of $^{10}$B$\MakeLowercase{e}$}
\label{sec:results}
The Monte Carlo method is used because of its superiority in
calculating the numerical integrations in the Hamiltonian kernels,
which otherwise is very difficult to be solved analytically for the
THSR wave functions of $^{10}$Be. The Monte Carlo technique is very
flexible for extending the THSR concept. The Monte Carlo calculation
includes the integrations of Euler angle $\Omega$ in the angular
momentum projection and integrations of generate coordinates
$\{\mathbf{R}, \mathbf{R}_n\}$ in the creation operators. When using
one single THSR wave function, the numerical calculation would be much
more efficient than traditional GCM calculations. To compare our
results with other models, the width of the Gaussians in the single
nucleon wave function is chosen to be $b=1.46$ fm, which is the same
as fixed in Refs.~\cite{Suhara2010, Kobayashi2012}. The $\beta$
parameters in the THSR wave functions are treated as variational
parameters and are optimized with the variational technique.

We investigate the ground state $0_{1}^{+}$ of $^{10}$Be and its
rotational band with both the independent THSR wave function and the
correlated THSR wave function of $^{10}$Be.  We choose the parameter
$m=1$ with spin up for one valence neutron and parameter $m=-1$ with
spin down for the other. This is to ensure parallel coupling of spin
and the orbital angular momentum for both neutrons as we used in
previous investigations \cite{Lyu2015}.  The optimum variational
parameters for the independent THSR wave function are
$\beta_{\alpha,xy}=0.1$ fm, $\beta_{\alpha,z}=2.0$ fm,
$\beta_{n,xy}=2.0$ fm and $\beta_{n,z}=3.5$ fm.  For the correlated
THSR wave function, the optimum variational parameters are
$\beta_{\alpha,xy}=0.1$ fm, $\beta_{\alpha,z}=2.2$ fm,
$\beta_{{\textrm{pair}},xy}=0.8$ fm, $\beta_{{\textrm{pair}}, z}=1.8$ fm,
$\beta_{n,xy}=2.0$ fm and $\beta_{n,z}=2.7$ fm.

In Table \ref{table:10base}, we list the calculated results of the
$0^{+}$ ground state of $^{10}$Be together with results from other
models and experimental values. The binding energies obtained for the
ground state $0_{1}^{+}$ are $-58.0$ MeV and $-58.2$ MeV with
independent and correlated THSR wave functions, respectively.
Theoretical calculations with other models, in which the same
potential is used, provide binding energies of the ground state of
$^{10}$Be ranging from about $-59$ MeV to $-60$ MeV, which is about
1$\sim$2 MeV lower than our results \cite{Kobayashi2011,
  Suhara2010}. This difference is reasonable because we are using only
one single THSR wave function what may not be the optimal choice for
the valence neutrons. If we superpose 8 THSR wave functions with
different parameters $\beta$ for $\alpha$-clusters or valence neutrons
in the THSR wave function, the ground state energy would decrease to
$-59.0$ MeV, which is consistent with other methods, as shown in Table
\ref{table:10base}. The experimental value for the ground state of
$^{10}$Be is $-65.0$ MeV, which is much lower than any theoretical
results from our and other groups' works. These differences originate
from the choice of effective potentials.

From these results, we can also notice that the binding energy of the
ground state is improved by about $0.2$ MeV by the introduction of the
valence neutron correlation in the THSR wave function. This
improvement shows the correlation effect of the valence
neutrons in the ground state, as we will discuss later in Section V.

In our present calculation, the ansatz of Gaussians (multiplied by
factors) is used for the valence neutron in the THSR wave functions. In
the future, we will discuss more general form of the THSR wave
function for a better optimized description of the valence neutrons.

\begin{table}[htbp]
\begin{center}
  \caption{\label{table:10base}Results obtained for the $0^{+}$
    ground state of $^{10}$Be. THSR$_{\textrm{ind}}$ and THSR$_{\textrm{cor}}$ are
    binding energies calculated with the independent THSR wave
    function and the correlated THSR wave function, respectively.
    THSR$_{\Sigma 8}$ denotes calculated result by superposing 8
    different THSR wave functions. Results from other theoretical
    methods are also listed.}
\begin{tabular}{l l l  l l}
\hline
\hline
Model &E (MeV)\\
\hline
THSR$_{\textrm{ind}}$ &-58.0 \\
THSR$_{\textrm{cor}}$ &-58.2 \\
THSR$_{\Sigma 8}$ &-59.0 \\
AMD\cite{Kobayashi2011} &-58.7 \\
AMD+DC\cite{Kobayashi2011} &-60.4 \\
AMD+GCM\cite{Suhara2010}  &-59.2 \\
\hline
\hline
\end{tabular}
\end{center}
\end{table}

\begin{figure}[htbp]
  \centering
  \includegraphics[width=0.45\textwidth]{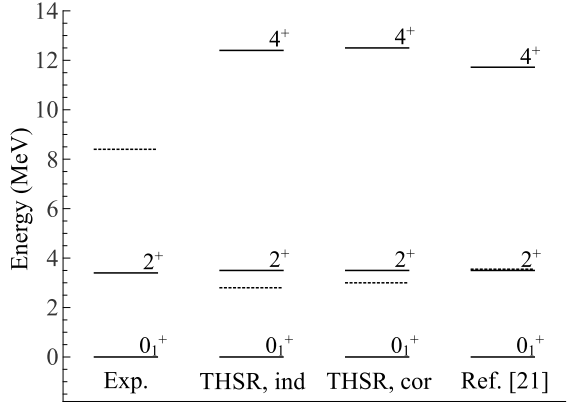}
  \caption{\label{fig:GS-Band}The $0^{+}$ ground state of $^{10}$Be
    and its rotational band. ``THSR,ind" and ``THSR,cor" denote
    calculated results with the independent THSR wave function and
    the correlated THSR wave function, respectively. ``Ref.~[21]" denotes the
     results of the AMD method ~\cite{Kobayashi2012}. ``Exp." denotes the experimental
    result. The dashed lines indicate the corresponding $\alpha+\alpha+n+n$
    threshold -55.2 MeV.}
\end{figure}

We show in Fig.~\ref{fig:GS-Band} the energy spectrum of the
$0_{1}^{+}$ rotational band of $^{10}$Be based on the ground state.
The calculated excitation energy of the $2_{1}^{+}$ state is 3.5 MeV and fits
very well with the experimental value 3.4 MeV.  Good agreement can
also be seen between our calculation and the AMD method for the
$2_{1}^{+}$ and $4_{1}^{+}$ excited states.

The  root-mean-square radius of the $0_{1}^{+}$ ground state of
$^{10}$Be is also obtained from our approach. The result is 2.57 fm with
one single THSR wave function, which is consistent with the value 2.5
fm from Ref.~\cite{Kobayashi2012}, but slightly larger than the value
2.37 fm from Ref.~\cite{Kobayashi2011} and the experimental value 2.30
fm. This small difference originates from the slightly weaker binding
effect described by only one single THSR wave function, as discussed above.

\begin{figure}[htbp]
  \centering
  \includegraphics[width=0.45\textwidth]{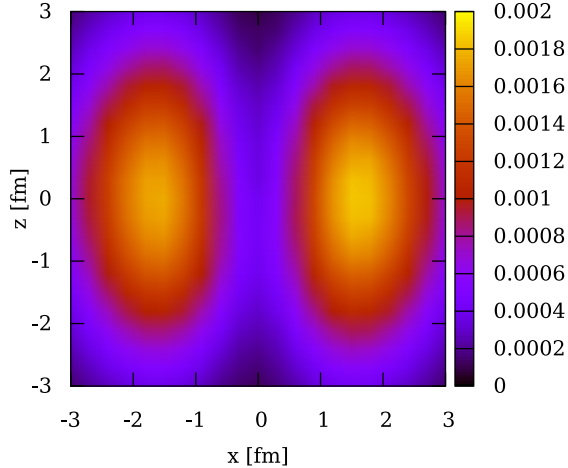}
  \caption{\label{fig:pi-dens}(Color online.) Density distribution of the valence
      neutrons in the intrinsic ground state of ${}^{10}$Be. The
    color scale of each point in the figure is proportional to the
    nucleon density on $x-z$ plane of the $y=0$ cross section. The
    unit of the density is fm$^{-3}$.  }
\end{figure}

The density distribution $\rho(\mathbf{r}'_n)$ of the extra nucleons
is calculated to give a clear view of the dynamics of the valence
neutrons. The intrinsic wave function $|\Psi\rangle$ of $^{10}$Be can
be written in the following form
\begin{equation}
  |\Psi\rangle = \text{C} {\mathcal A}[\Phi^{ \text{THSR}}(2 \alpha)
     \phi_{{\textrm{pair}}}(\mathbf{r}_{1},\mathbf{r}_{2})],
\end{equation}
where ${\mathcal A}$ is the antisymmetrizer and C is a normalization
constant.  Then the density distribution $\rho(\mathbf{r}')$ of the
valence neutrons is defined as
\begin{equation}
\label{eq:density}
  \rho(\mathbf{r}') = N_{c} \langle \Phi^{ \text{THSR}}(2 \alpha)
    \phi_{{\textrm{pair}}}(\mathbf{r}_{1},\mathbf{r}_{2})|
    \delta(\mathbf{r}_{1} - \mathbf{X}_G - \mathbf{r}')+
    \delta(\mathbf{r}_{2} - \mathbf{X}_G - \mathbf{r}')
  | \Psi \rangle,
\end{equation}
where $N_{c}$ is the normalization constant \cite{Horiuchi1977}.  As
shown in Fig.~\ref{fig:pi-dens}, the density distribution of two
valence neutrons has the same shape as the one which we obtained for the ground
state $^{9}$Be \cite{Lyu2015} with only a single valence neutron. The extension of the density distribution
in the $z$-direction and the absence of neutrons along the
$z$-axis shows a good description of the $\pi$-orbit in the ground state
of $^{10}$Be as suggested by nuclear molecular orbit (MO) model
\cite{Oertzen1996, Itagaki2000}. This reproduction of $\pi$-orbit
structure is obtained naturally from the antisymmetrization in the
THSR wave function which cancels non-physical distribution of valence
neutrons, e.g. positions close to the center of $\alpha$-clusters.

\section{The $0_{2}^{+}$ chain state of $^{10}$B\MakeLowercase{e}}
\begin{figure}[htbp]
  \centering
  \includegraphics[width=0.45\textwidth]{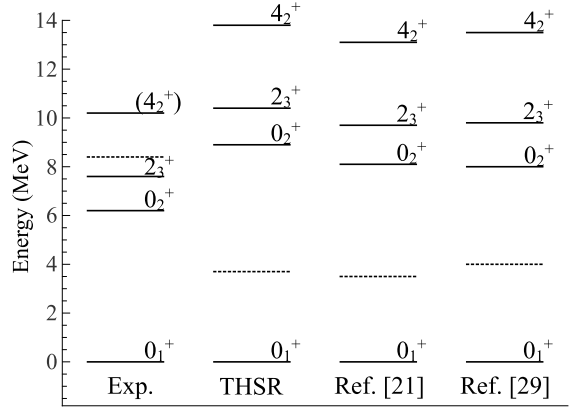}
  \caption{\label{fig:ES-Band} Energy spectrum of the $0_{2}^{+}$
    rotational band relative to the ground state energy.  The one
    labeled with Ref.~[21] is the value calculated with the AMD+DC
    method ~[21].  The one labeled with Ref.~[29] is the value
    calculated with the $\beta-\gamma$ constrained AMD+GCM method ~[29]. The dashed lines are the corresponding
    $\alpha+\alpha+n+n$ thresholds of -55.2 MeV.}
\end{figure}

In this section we study the $0_{2}^{+}$ state of $^{10}$Be with one
single THSR wave function as shown in Eq.~(\ref{eq:chainWF}).  The
optimum parameter $D$ in Eq.~(\ref{eq:extraCreatorSigma}) is $D=2.0$
fm.  Other optimum variational parameters in the THSR wave function
are $\beta_{\alpha}=3.5$ fm and $\beta_{n}=4.0$ fm.  The calculated
spectrum for the $0_{2}^{+}$ rotational band based on the ground state
is shown in Fig.~\ref{fig:ES-Band}. The ground state energy from THSR
calculation in this figure is chosen to be the one with -59.0 MeV
obtained from the superposed THSR wave functions as listed in Table
\ref{table:10base}. Fig.~\ref{fig:ES-Band} shows systematical
discrepancies between theoretical results calculated by different
models and the experimental values. This is because of the choice of
effective interactions. The calculated energy spectrum of the
$0_{2}^{+}$ rotational band with the THSR wave function, as shown in
Fig.~\ref{fig:ES-Band}, agrees well with results of other theoretical
models from Refs.~\cite{Suhara2010, Kobayashi2012}. The THSR wave
function also gives energy gaps between the states in the $0_{2}^{+}$
band which fit very well with experimental data and other models. It
is very interesting to see that one single THSR wave function can
describe well the $0_{2}^{+}$ state of $^{10}$Be, while in other
theoretical models superposition of large number of basis sets is
needed.

We checked the orthogonality between the THSR wave function for the
$0_{2}^{+}$ state and  the $0_{1}^{+}$ ground state.  The
calculated overlap between these two states is 1.4\%, which satisfies
the requirement for eigenstates.

We also calculate the root-mean-square radius for the
$0_{2}^{+}$ state of $^{10}$Be. The calculated result is 3.11 fm,
which is consistent with 2.96 fm from Ref.~\cite{Kobayashi2011} and
3.4 fm from Ref.~\cite{Kobayashi2012}.

\begin{figure}[htbp]
  \centering
  \includegraphics[width=0.3\textwidth]{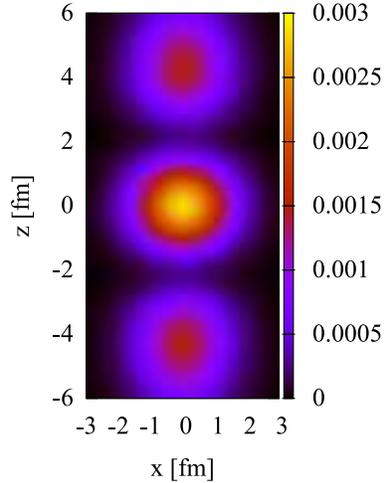}
  \caption{\label{fig:sigma-dens}(Color online.) Density distribution of valence
    neutrons in the intrinsic $0_{2}^{+}$ state of ${}^{10}$Be. The
    color scale of each point in the figure is proportional to the
    nucleon density in the $x-z$ plane of the $y=0$ cross section. The
    unit of the density is fm$^{-3}$.  }
\end{figure}

To illustrate the structure of the $0_{2}^{+}$ state of $^{10}$Be with the
THSR wave function, we show the density distribution
$\rho(\mathbf{r}'_n)$ of valence nucleons in
Fig.~\ref{fig:sigma-dens}.  It is clearly seen that the distribution
of valence neutrons is divided into three regions separated by two
nodes perpendicular to the $z$-axis, which is a typical character of
the $\sigma$-orbit. The node structure originates from the newly
introduced factor $(D-|R_{n,z}|)$ in
Eq.~(\ref{eq:extraCreatorSigma}). Also a large spread of more than 10 fm
along the $z$-axis is observed for the valence neutrons, which is one of
the reasons for the enormously large spatial extension of the $0_{2}^{+}$
state in $^{10}$Be. Another reason is the large $\alpha$-cluster
distribution as discussed in the next section.

\section{Analysis of cluster dynamics and dineutron correlations}
In this section, we discuss the dynamics of the $\alpha$-clusters and
the valence neutrons. The THSR wave function is based on the
nonlocalized concept of cluster dynamics which describes the
$\alpha$-cluster motion by a Gaussian style THSR integration of
generator coordinates as in Eq.~(\ref{eq:alphaCreator}) or
Eq.~(\ref{eq:alphaCreatorSigma}). 
The optimum $\beta$-parameters in the $\alpha$-creation operators is
$\beta_{\alpha,xy}\approx 0$ fm, $\beta_{\alpha,z}=2.0$ fm for the
ground $0_{1}^{+}$ state while the optimum value
$\beta_{\alpha}=3.5$ fm is obtained for the $0_{2}^{+}$
state. Considering that large $\beta$ parameters corresponds to large
extension of the nonlocalized cluster motion, it is clear that the
$\alpha$-particles are much more tightly bound by the $\pi-$orbit than by
the $\sigma$-orbit. The $\pi$-binding effect in the ground state of
$^{10}$Be is also stronger than that in the ground state of $^{9}$Be,
where the optimum parameter for $\alpha$-cluster is
$\beta_{\alpha,z}=4.2$ fm \cite{Lyu2015}.

\begin{figure}[htbp]
  \centering
  \includegraphics[width=0.45\textwidth]{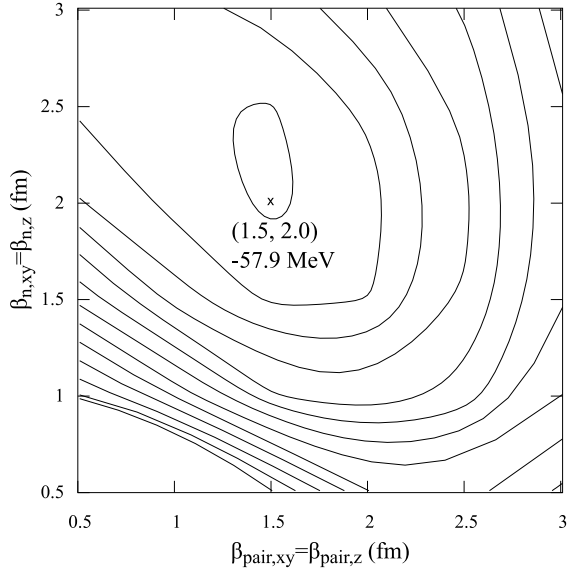}
  \caption{\label{fig:cor-contour}Contour map of the binding energy
    surface of the ground state with different $\beta$ parameters in
    the correlated THSR wave function.  The horizontal coordinates are
    the $\beta_{{\textrm{pair}}}$ parameters for the center-of-mass of two
    valence neutrons as $\beta_{{\textrm{pair}},xy}=\beta_{{\textrm{pair}},z}$.  The
    vertical coordinates are the $\beta_{n}$ parameters for each
    neutron in the two neutron sub-system as
    $\beta_{n,xy}=\beta_{n,z}$. Other parameters are
    $\beta_{\alpha,xy}=0.1$ fm,$\beta_{\alpha,z}=2.0$ fm.  The optimum
    value is marked on the map labeled with corresponding
    coordinates. }
\end{figure}

Another interesting problem is the dineutron correlation effect in the
$^{10}$Be nucleus.  With the correlated THSR wave function, we
calculate the contour map of the binding energy surface for the ground
state of $^{10}$Be as shown in Fig.~\ref{fig:cor-contour}.  For
simplicity, we fix the deformation between $x,y$ and $z$ directions to
be 0.  The optimum binding energy locates near the coordinates $(1.5,2.0)$,
where $\beta_{{\textrm{pair}},xy}=\beta_{{\textrm{pair}},z}=1.5$ fm and
$\beta_{n,xy}=\beta_{n,z}=2.0$ fm. This optimum value is slightly
higher than the final result of -58.2 MeV because deformation is
neglected here. When $\beta_{{\textrm{pair}}}=0$, the center of the two-neutron
sub-system is fixed at the origin of coordinates, and the correlated
THSR wave function turns to become an independent one for the two neutrons. Thus, the
optimum value of parameter $\beta_{n}>0$ shows the existence of 
  correlation effects in the ground state, as discussed above. Large
distance between the two neutrons in the dineutron pair can be concluded because of the
big value of the parameter $\beta_{n}$.  The relatively small value of the parameter
  $\beta_{{\textrm{pair}}}$ describes the collective motion of the two-neutron
sub-system. We also study the correlation of the two valence neutrons in
the $0_{2}^{+}$ state of $^{10}$Be. A very large value of the parameter
$\beta_{n}$ is obtained which shows nearly independent motion of the two
valence neutrons in this state.

\section{Conclusion}
\label{sec:conclusion}
We investigated the N=Z+2 nucleus $^{10}$Be from the nonlocalized
clustering concept. THSR wave functions with $\pi$-orbit structure and
chain ($\sigma$-orbit) structure are formulated for $^{10}$Be.
Correct parity and node structures are ensured in THSR wave functions
for the corresponding states. The $0_{1}^{+}$ ground state of $^{10}$Be
and its rotational band is calculated and the results agree well with
other models and experimental values.  Small improvement is observed
for the binding energy of the ground state with introduction of
correlations between two valence neutrons in a single one THSR wave
function. The $0_{2}^{+}$ state is also studied with the newly
formulated THSR wave function containing a node factor. The
calculated energy spectrum of the $0_{2}^{+}$ rotational band is
consistent with values from other models as well as with
experiments. This result is very interesting because only one single
THSR wave function is used.  Root-mean-square radii are also
calculated for the first two $0^{+}$ states of $^{10}$Be, which have
good agreements with other models.  The density distribution of
valence neutrons shows good description of $\sigma$-orbit by the THSR
wave function. It is the first application of the nonlocalized picture
to $\sigma$-orbit binding systems. Analysis of optimum $\beta$
parameters shows much tighter binding effect for $\alpha$-clusters
within the $\pi$-orbit structure than with the $\sigma$-orbit
structure.  Showing the contour map of energy calculated with
correlated THSR wave function, we discussed the correlation effect
between valence neutrons and also the intrinsic and collective motion
of the di-neutron pair in the ground state. The investigation of
$^{10}$Be is another extension of the nonlocalized concept and of the
THSR wave function towards more general nuclear structures.  Our
calculations with the THSR wave function and the Monte Carlo technique
requires less numerical work than the traditional GCM treatment. Also, the
THSR wave function used in this work illustrates more
physical insights of the $^{10}$Be nucleus.  In the future, this
scheme based on nonlocalized concept is also promising for the study
of neutron-rich nuclei with cluster structures and the investigations
of their corresponding cluster and nucleon dynamics.

\begin{acknowledgments}
  The authors would like to thank Professor Kanada-En'yo-san for valuable
  discussions.  This work is supported by the National Natural Science
  Foundation of China (grant nos 11535004, 11375086, 11120101005,
  11175085 and 11235001, 11575082), by the 973 National Major State
  Basic Research and Development of China, grant nos 2013CB834400 and
  by the Science and Technology Development Fund of Macao under grant
  no. 068/2011/A.
\end{acknowledgments}

\end{document}